%% file: main.tex
\documentclass[10pt]{article}
\usepackage[preprint]{tmlr}

\usepackage{amsmath,amssymb}
\usepackage{graphicx}
\usepackage{xcolor}
\usepackage{booktabs}
\usepackage{hyperref}
\usepackage{cleveref}
\usepackage{algorithm}
\usepackage{algpseudocode}
\usepackage{amsthm}
\usepackage{multirow}
\usepackage{subfig}

\newcommand{\like}{\mathcal{L}}
\newcommand{\loglike}{\ell}
\newcommand{\data}{\mathcal{D}}
\newcommand{\dkl}{D_{\mathrm{KL}}}
\newcommand{\prior}{\pi}
\newcommand{\posterior}{p}
\newcommand{\evidence}{\mathcal{Z}}
\newcommand{\hyper}{\psi}
\newcommand{\local}{\theta}
\newcommand{\thresh}{\loglike^*}
\newcommand{\budget}{B}
\newcommand{\bigO}{\mathcal{O}}

\title{Nested Sampling with Slice-within-Gibbs: Efficient Evidence Calculation for Hierarchical Bayesian Models}

\author{\name David Yallup \email dy297@cam.ac.uk \\
      \addr Kavli Institute for Cosmology Cambridge, University of Cambridge\\
      Institute of Astronomy, University of Cambridge}

\def\month{02}
\def\year{2026}
\def\openreview{\url{https://openreview.net/forum?id=XXXX}}

\begin{document}

\maketitle

\begin{abstract}
  We present Nested Sampling with Slice-within-Gibbs (NS-SwiG), an algorithm for Bayesian inference and evidence estimation in high-dimensional models whose likelihood admits a factorization, such as hierarchical Bayesian models. We construct a procedure to sample from the likelihood-constrained prior using a Slice-within-Gibbs kernel: an outer update of hyperparameters followed by inner block updates over local parameters. A likelihood-budget decomposition caches per-block contributions so that each local update checks feasibility in constant time rather than recomputing the global constraint at linearly growing cost. This reduces the per-replacement cost from quadratic to linear in the number of groups, and the overall algorithmic complexity from cubic to quadratic under standard assumptions. The decomposition extends naturally beyond independent observations, and we demonstrate this on Markov-structured latent variables. We evaluate NS-SwiG on challenging benchmarks, demonstrating scalability to thousands of dimensions and accurate evidence estimates even on posterior geometries where state-of-the-art gradient-based samplers can struggle.
\end{abstract}

\input{sections/introduction}

\input{sections/background}

\input{sections/method}


\input{sections/experiments}

\input{sections/discussion}

\input{sections/conclusion}

\subsubsection*{Acknowledgments}
The author thanks Will Handley for useful discussions. The author was supported by the research environment and infrastructure of the Handley Lab at the University of Cambridge. The author used OpenAI GPT 5.2 to refine portions of the draft, and Claude Opus 4.5 was used in refining the code. The author takes full responsibility for the final content.

\subsubsection*{Code and Data Availability}
The algorithm code and example scripts to reproduce the experiments in this paper are available at \url{https://github.com/yallup/swig}

\bibliographystyle{tmlr}
\bibliography{references}

\input{appendix}

\end{document}

%% file: sections/introduction.tex
\section{Introduction}
\label{sec:intro}

Nested sampling \citep{Skilling2006} has become a cornerstone of Bayesian inference in the physical sciences \citep{Ashton2022,Buchner2023}.
Unlike Markov chain Monte Carlo (MCMC), nested sampling directly estimates the Bayesian evidence $\evidence = \int \like(\data|\local)\, \prior(\local)\, d\local$, enabling rigorous model comparison.
This is central to hypothesis testing tasks such as comparing physical models, distinguishing signal from noise, or selecting among population models.
Practical implementations treat constrained-prior sampling---drawing from the prior truncated to a likelihood threshold---largely as a black-box task~\citep{Feroz2009,Handley2015,Speagle2020,buchner2021ultranestrobustgeneral}, with limited exploitation of model-specific factorisation.
In high dimensions, the available constrained MCMC mutation kernels can mix poorly, often resembling local random-walk exploration.

The dominant paradigm for scaling Bayesian methods to high dimensions is to use the gradient of the target distribution to suppress this random-walk behaviour~\citep{fearnhead2024scalablemontecarlobayesian}.
Gradient-based samplers such as Hamiltonian Monte Carlo (HMC) \citep{Neal2011} and the No-U-Turn Sampler (NUTS) \citep{Hoffman2014} have become ubiquitous general-purpose inference tools, and the success of Stan~\citep{Carpenter2017} has reinforced a widespread view---particularly among practitioners---that gradients are essential for scalability.
Designing efficient gradient-based samplers for hard likelihood constraints remains challenging; existing approaches for nested sampling~\citep{Feroz_2013,lemos2023improvinggradientguidednestedsampling} have not yet demonstrated robust performance across generic problem classes~\citep{kroupa2025resonances}, outside settings with additional structure (e.g., log-concave models amenable to proximal methods~\citep{Cai_2022}).
This has left nested sampling viewed as well suited for low-dimensional problems with pathological geometries, but requiring alternatives for high-dimensional inference~\citep{Piras_2024}.

This paper introduces \emph{Nested Sampling with Slice-within-Gibbs} (NS-SwiG), a constrained-sampling inner loop that exploits the conditional independence structure of models with factorised likelihoods.
Consider a hierarchical model with $J$ groups, local parameters $\local_j$, and shared hyperparameters $\hyper$, where the log-likelihood decomposes as $\loglike(\hyper, \{\local_j\}) = \sum_{j=1}^J \loglike_j(\local_j, \hyper)$.
Standard nested sampling checks the global constraint $\loglike > \thresh$ at every MCMC proposal, costing $\bigO(J)$ per evaluation; with $\bigO(J)$ parameters requiring updates, this yields $\bigO(J^2)$ per replacement.
NS-SwiG decomposes the global constraint into per-block budgets checkable in $\bigO(1)$, reducing the cost to $\bigO(J)$.

This builds on contemporary work reviving interest in coordinate-wise \emph{Metropolis-within-Gibbs} schemes as scalable alternatives to joint-gradient methods~\citep{Ascolani2024,Luu2024}.
We embed a Slice-within-Gibbs kernel within a nested sampling outer loop, targeting the joint constrained prior with block updates: an outer slice update of hyperparameters followed by an inner sweep over the $J$ local blocks.
We demonstrate the scalability this unlocks on challenging benchmarks up to $d \sim 2500$ dimensions. In summary, we make the following contributions:
\begin{enumerate}
\item[\textit{(i)}] Likelihood-budget decomposition. We show that the global nested sampling constraint can be decomposed into per-block budgets updated in $\bigO(1)$ from a cached total, extending naturally to Markov-structured latent variables where budgets depend on local neighbourhoods.
\item[\textit{(ii)}] Constrained Slice-within-Gibbs kernel. We construct an MCMC kernel for the joint constrained prior using blocked slice updates, yielding $\bigO(J)$ per sweep compared to $\bigO(J^2)$ for standard joint-space constrained sampling.
\item[\textit{(iii)}] Implementation and validation. We provide a fully vectorised JAX implementation building on~\citet{Yallup2026}, validate posterior recovery against NUTS on four benchmarks, and demonstrate accurate evidence estimates at scale. We motivate a number of immediate applications within astrophysics, where nested sampling is already widely used for object-level inference problems.
\end{enumerate}

The remainder of the paper is organized as follows.
Section~\ref{sec:background} reviews nested sampling fundamentals and hierarchical model structure.
Section~\ref{sec:method} presents NS-SwiG, including the budget decomposition, caching strategy, and constrained Slice-within-Gibbs updates.
Section~\ref{sec:experiments} reports numerical experiments validating correctness and scaling across four benchmarks.
Section~\ref{sec:discussion} discusses limitations, extensions, and practical considerations, and Section~\ref{sec:conclusion} concludes.

%% file: sections/background.tex
\section{Background}
\label{sec:background}

Many high-dimensional Bayesian models are high-dimensional because they compose many repeated low-dimensional components. Consider an inference task where $J$ objects are observed with data $\data_j$; we construct a model with local parameters $\local_j$ and shared hyperparameters $\hyper$. The likelihood factorizes as
\begin{equation}
\like(\data \mid \hyper, \{\local_j\}_{j=1}^J) = \prod_{j=1}^J \like_j(\data_j \mid \local_j, \hyper).
\label{eq:hierarchical_likelihood}
\end{equation}
Defining $\loglike_j(\local_j, \hyper) := \log p(\data_j \mid \local_j, \hyper)$, the log-likelihood decomposes as $\loglike(\hyper, \{\local_j\}) = \sum_{j=1}^J \loglike_j(\local_j, \hyper)$.

While this is a restricted class of models, it is ubiquitous in modern astrophysics and cosmology \citep{shariff_bahamas_2016,2016MNRAS.460.4258L,Mandel2019}. Such models arise naturally when analyzing populations of objects---gravitational-wave sources \citep{Thrane2019,Abbott2023}, supernovae~\citep{grayling2024scalablehierarchicalbayesninference}, galaxies~\citep{2024ApJS..274...12A}---where each observation constrains local parameters while shared hyperparameters describe the population-level distribution. In many of these examples, object-level posteriors can be multimodal or have invalid regions of parameter space, making nested sampling a practical solution~\citep{bilby_paper,prospector}. However, this can make population-level inference challenging: lifting object-level analyses into a joint gradient-based sampler often compromises the robustness of the object-level inference. 

This provides domain-specific motivation, but more generally, extracting marginal likelihoods for high-dimensional models is broadly relevant across scientific domains and remains notoriously challenging~\citep{mlreview}. In this section we review the necessary background on nested sampling, Metropolis-within-Gibbs, and hierarchical models.

\subsection{Nested Sampling}
\label{subsec:ns}

Let the full parameter vector be $\vartheta \equiv (\hyper, \local_{1:J}) \in \mathbb{R}^d$ with $d = d_\hyper + J d_\local$, prior density $\prior(\vartheta)$, likelihood $\like(\vartheta) = p(\data \mid \vartheta)$, log-likelihood $\loglike(\vartheta) = \log \like(\vartheta)$, and evidence
\begin{equation}\label{eq:evidence}
\evidence = \int \like(\vartheta)\,\prior(\vartheta)\,d\vartheta.
\end{equation}
Nested sampling \citep{Skilling2006} rewrites this integral using the prior volume above a log-likelihood threshold,
\begin{equation}
X(\ell) \;:=\; \int \prior(\vartheta)\,\mathbf{1}_{\{\loglike(\vartheta) > \ell\}}\,d\vartheta \;\in\; [0,1],
\label{eq:prior_volume}
\end{equation}
and let $\ell(X)$ denote its (generalized) inverse.
Then $\evidence = \int_0^1 \like(X)\,dX$, where $\like(X) = \exp(\ell(X))$ is the likelihood at prior volume $X$.

The algorithm maintains $m$ particles (often called \emph{live points} in the nested sampling literature) approximating draws from the \emph{constrained prior} at the current threshold $\thresh$.
At each iteration it removes the live point with the smallest log-likelihood, accounts for its contribution to $\evidence$, and replaces it with a new draw from
\begin{equation}
\prior_{\thresh}(\vartheta) \;\propto\; \prior(\vartheta)\,\mathbf{1}_{\{\loglike(\vartheta) > \thresh\}},
\label{eq:constrained_prior}
\end{equation}
with $X(\thresh)$ as the normalizing constant.
Efficiently sampling from \eqref{eq:constrained_prior} is typically the dominant cost, addressed by ellipsoidal methods \citep{Feroz2009}, slice-based approaches \citep{Handley2015,Neal2003}.

Nested sampling is widely used in the physical sciences for Bayesian model comparison via marginal likelihood (evidence) estimation, often in settings with computationally expensive forward models. From this perspective it is closely related to Sequential Monte Carlo (SMC) samplers for static targets, which estimate normalizing constants using an artificial sequence of intermediate distributions~\citep{chopin_sequential_2002,DelMoral2006}, often defined by likelihood tempering/annealing~\citep{neal2001annealedimportancesampling}. We do not address popular scalable approaches to model comparison based on stacking or cross-validation~\citep{vehtari2017practical,fong2019marginal,Yao_2022}. These methods primarily address scalability with respect to large datasets (e.g., avoiding refits for held-out data) and typically operate by post-processing draws from an existing inference procedure; they are therefore not themselves a sampling technique.

\subsection{Metropolis-within-Gibbs and Hierarchical Models}
\label{subsec:mwG}
Gibbs sampling is a classic algorithm for sampling from a joint distribution by iteratively sampling from the full conditional of each variable given the others~\citep{mackay}, often employed in situations where the full conditionals are tractable and easy to sample from. Metropolis-within-Gibbs generalizes the idea of moving a subset of parameters whilst fixing others~\citep{chib_understanding_1995}. For challenging conditional distributions, slice sampling~\citep{Neal2003} within Gibbs-style alternating updates has been shown to mix effectively even in high dimension for generalized linear models~\citep{Luu2024}.

This aligns well with two aspects of our setting. First, slice sampling is the dominant paradigm used for MCMC mutation within nested sampling~\citep{Handley2015}. Second, hierarchical models are a classic setting for Gibbs sampling~\citep{damien_gibbs_1999}, though practical implementation for general non-conjugate models and large $J$ remains challenging.

For hierarchical models as in \cref{eq:hierarchical_likelihood}, with hyperparameters $\hyper \in \mathbb{R}^{d_\hyper}$ and group-specific parameters $\local_j \in \mathbb{R}^{d_\local}$ for $j = 1, \ldots, J$, we assume the prior factorizes as $\prior(\vartheta) = \prior(\hyper) \prod_j \prior(\local_j \mid \hyper)$.
The joint posterior is then
\begin{equation}
\posterior(\hyper, \local_{1:J} \mid \data) \propto \prior(\hyper) \prod_{j=1}^J \prior(\local_j \mid \hyper)\,\like_j(\data_j \mid \local_j, \hyper).
\label{eq:hierarchical_posterior}
\end{equation}
This structure motivates sampling \eqref{eq:constrained_prior} using block Gibbs updates of $\hyper$ and each $\local_j$ within the constrained sampler.
The key scaling issue is the likelihood constraint check $\loglike(\vartheta) > \thresh$: a naive implementation recomputes \eqref{eq:hierarchical_likelihood} from scratch after each block proposal, costing $\bigO(J)$ per proposal.
Over a sweep that updates all $J$ local blocks, this yields $\bigO(J^2)$ work per sweep.
By caching the current per-group likelihood contributions $\{\loglike_j\}_{j=1}^J$ and their sum, a proposal that changes only one block can update the constraint in $\bigO(1)$, reducing the sweep cost to $\bigO(J)$. Additional savings arise when the data likelihood $p(\data_j | \local_j)$ does not depend on $\hyper$: only the conditional prior $\prior(\local_j | \hyper)$ needs recomputation during hyperparameter updates, and this is typically much cheaper than the data likelihood.
Kernels of this \emph{within-Gibbs} type leave the full joint target invariant (up to MCMC convergence), and recent results establish dimension-robust convergence for Metropolis-within-Gibbs on broad classes of hierarchical models \citep{Ascolani2024}.

In standard nested sampling, sampling from the constrained prior typically requires MCMC moves to decorrelate the new sample from its parent. A standard heuristic to take is that adequate decorrelation in practice typically requires $\bigO(d)$ slice sampling steps, where $d$ is the total parameter dimension~\citep{Yallup2026}. For hierarchical models, $d = d_\hyper + J \cdot d_\local$, where $d_\hyper$ and $d_\local$ are the hyperparameter and per-group local parameter dimensions respectively.
Each slice step must check the likelihood constraint, which in the naive implementation requires computing the full sum above at cost $\bigO(J)$.
This yields a per-replacement cost of $\bigO(d \cdot J) = \bigO(J^2)$ for large $J$, which becomes prohibitive for object catalogs with hundreds to thousands of groups.

High dimension further exacerbates some issues in constrained slice sampling.
Convergence bounds for Hit-and-Run within slice sampling exhibit polynomial dependence on dimension, with spectral gaps scaling as $\bigO(d^{-3})$ for log-concave targets \citep{Power2024}.
Moreover, a practical strength of nested sampling---its weak reliance on hand-tuned proposals---depends on representing the geometry of the constrained region using the live point cloud.
As $d$ grows, both the MCMC exploration and this geometric adaptation become less effective, compounding the computational bottleneck in hierarchical settings. Decomposing high-dimensional constrained sampling into lower-dimensional conditional updates can substantially improve both computational cost and mixing.

\subsection{Structure-Aware Inner Kernels in Nested Sampling}
\label{subsec:related}

Two lines of prior work exploit model structure within the nested sampling inner loop. The first is the \emph{fast--slow} hierarchy of parameters, developed for cosmological likelihoods~\citep{Lewis:2013hha} and implemented in PolyChord \citep{Handley2015,Handley2015b}. When the likelihood factors as $\like(z,\theta) = \like(f(\psi), \theta)$ for an expensive forward model $z = f(\psi)$, the inner MCMC can propose moves in the ``fast'' parameters $\theta$ more frequently and cache the forward-model evaluations. This exploits heterogeneous evaluation costs but does not decompose the likelihood constraint itself. Secondly, \citet{NIPS2005_9dc37271} introduced block Gibbs updates within nested sampling for the Potts model, using Swendsen--Wang~\citep{swendsen_nonuniversal_1987} cluster moves as the constrained inner kernel. The random-cluster auxiliary-variable augmentation yields tractable conditionals on a lattice, but the construction is specific to discrete Markov random fields and does not transfer to general continuous hierarchical models. 

Both approaches can be viewed as special cases of a common model structure-aware principle: design the nested sampling mutation kernel compositionally, using blocked updates paired with inner moves that exploit model-specific structure. We show that this compositional perspective can be made broadly effective at scale by using a general-purpose Slice-within-Gibbs constrained sampler as the building block. In the hierarchical setting we emphasize, the additive log-likelihood further enables a clean transformation of the global likelihood constraint into per-block budgets that can be maintained and checked in constant time. We also note that there are many examples of custom MCMC moves in nested sampling targeting specific models, such as the reversible jump~\citep{green_reversible_1995} moves of DNest~\citep{Brewer:2016scw}. These are mostly positioned as improving the mixing of the inner sampler---certainly an intended side effect of the compositional perspective we advocate---but do not explicitly target cost and scalability.

%% file: sections/method.tex
\section{Method and Implementation}
\label{sec:method}
With the tools from the previous section established, we present the NS-SwiG algorithm, and discuss its implementation details.

\subsection{Nested Sampling Budget Constraint Decomposition}
\label{subsec:budget}

As outlined in \cref{eq:hierarchical_likelihood}, we consider a hierarchical Bayesian model with hyperparameters $\hyper \in \mathbb{R}^{d_\hyper}$ and $J$ groups, each with local parameters $\local_j \in \mathbb{R}^{d_\local}$.
The per-group log-likelihood is $\loglike_j(\local_j, \hyper)$, and the total log-likelihood is the sum $\loglike = \sum_{j=1}^J \loglike_j$.

The central insight of NS-SwiG is that the nested sampling constraint can be decomposed per-group.
Fix $\hyper$ and $\{\local_j\}_{j \neq k}$, and consider proposing $\local_k'$ for group $k$. The global constraint $\sum_{j=1}^J \loglike_j > \thresh$ then rearranges to
\begin{equation}
\loglike_k(\local_k', \hyper) > \budget_k, \qquad \budget_k \equiv \thresh - S + \loglike_k(\local_k, \hyper),
\label{eq:budget}
\end{equation}
where $S = \sum_{j=1}^J \loglike_j(\local_j, \hyper)$ is the cached total log-likelihood at the current state. Equivalently, $\budget_k = \thresh - \sum_{j \neq k} \loglike_j(\local_j, \hyper)$.
Given $S$ and the current $\loglike_k$, computing the budget requires only $\bigO(1)$ operations, reducing constraint-checking from $\bigO(J)$ to $\bigO(1)$ per local update.%
\footnote{This decomposition can be viewed through the lens of auxiliary budget variables $\{B_j\}$ with $B_j \leq \loglike_j$ and $\sum_j B_j \geq \thresh$, so that the global constraint decomposes into per-factor bounds---analogous to the auxiliary-variable constructions of \citet{damien_gibbs_1999}.}

Similar residual-constraint decompositions appear in coordinate descent, Gibbs sampling for truncated distributions, and incremental likelihood caching \citep{Luu2024}; our contribution is embedding this idea within nested sampling's constrained prior, where caching per-group contributions $\{\loglike_j\}_{j=1}^J$ and their sum $S$ enables $\bigO(1)$ incremental updates upon acceptance of any local proposal. Since $\budget_k = \loglike_k - (S - \thresh)$, the budget equals the current group log-likelihood minus the global slack $(S - \thresh)$.

In practice this means we maintain two cached quantities: the running total $S$ and a vector of per-group log-likelihoods $\{\loglike_j\}_{j=1}^J$.
Upon acceptance of a proposal $\local_k'$, the algorithm computes $\loglike_k' = \loglike_k(\local_k', \hyper)$, updates $S \leftarrow S - \loglike_k + \loglike_k'$, and stores $\loglike_k'$---all in $\bigO(1)$.
A complete sweep costs $\bigO(J)$, a factor of $J$ improvement over recomputing the full sum at each step. Early in nested sampling, when $\thresh$ is low and slack is large, $\budget_k \ll \loglike_k$, giving group $k$ freedom to explore. As $\thresh$ increases and slack vanishes, $\budget_k \to \loglike_k$, forcing each group to concentrate in high-likelihood regions, the correct behavior for the nested sampling procedure.

More generally, if the log-likelihood can be written as a sum of local contributions, then when we update a block of variables we only need to recompute the terms that actually involve that block. The cost of checking the nested-sampling likelihood constraint for a proposed block move is therefore proportional to the number of affected terms, rather than to the full data set. This recovers the iid setting as the best case (each block touches only one term) and the Markov chain setting as the next simplest case (each block touches only a constant number of neighboring terms), which we work out in \cref{app:markov_swig}. In the worst case---when no useful decomposition is available---every update requires recomputing the full log-likelihood and we fall back to joint-space nested sampling.

\subsection{Constrained Slice-within-Gibbs Kernel}
\label{subsec:gibbs_theory}

With SwiG, we target sampling from the joint constrained prior
\begin{equation}
\prior_{\thresh}(\hyper, \{\local_j\}) \propto \prior(\hyper) \prod_{j=1}^J \prior(\local_j | \hyper) \cdot \mathbf{1}\left[\sum_j \loglike_j > \thresh\right]
\label{eq:joint_constrained_prior}
\end{equation}
via two alternating block updates, each implemented by slice sampling with stepping-out and shrinkage:

\paragraph{Hyperparameter update: $\hyper | \{\local_j\}$.}
Sample $\hyper$ from its full conditional
\begin{equation}
\prior_{\thresh}(\hyper | \{\local_j\}) \propto \prior(\hyper) \prod_{j=1}^J \prior(\local_j | \hyper) \cdot \mathbf{1}\left[\sum_j \loglike_j(\local_j, \hyper) > \thresh\right].
\label{eq:psi_conditional}
\end{equation}
The target includes \emph{both} the hyperprior and the conditional priors $\prod_j \prior(\local_j | \hyper)$; omitting the latter would target the wrong distribution. Each slice evaluation recomputes all $J$ group likelihoods, costing $\bigO(J)$. This step can be repeated $M_\hyper$ times per replacement, with the total cost scaling as $\bigO(M_\hyper \cdot J)$. For most reasonable hierarchical models the hyperparameter space is low-dimensional, so we follow the standard heuristic of fixing this to be equal to the number of hyperparameters, $M_\hyper = d_\hyper$.

\paragraph{Inner sweep: $\local_j | \hyper, \local_{-j}$ for $j = 1, \ldots, J$.}
For each group $j$, sample $\local_j$ from
\begin{equation}
\prior_{\thresh}(\local_j | \hyper, \local_{-j}) \propto \prior(\local_j | \hyper) \cdot \mathbf{1}\left[\loglike_j(\local_j, \hyper) > \budget_j\right],
\label{eq:theta_conditional}
\end{equation}
where the budget $\budget_j = \thresh - S + \loglike_j$ reformulates the global constraint as a per-group threshold.
Each slice evaluation requires a single group-likelihood call; since the number of slice evaluations per update has low variance and is insensitive to dimension under standard tuning~\citep{Yallup2026}, each local update costs $\bigO(1)$ in expectation. When $d_\local > 1$, each slice step proposes along a random direction drawn from the block-diagonal covariance estimate (hit-and-run), rather than cycling through coordinates. Since each local block is typically low-dimensional, we fix the number of inner slice sampling steps to $M_\local = d_\local$. This step comprises a sequential sweep through all $J$ groups, yielding a total cost of $\bigO(M_\local \cdot J)$. The overall cost of one Gibbs sweep is thus $\bigO((M_\hyper + M_\local) \cdot J)$, which for fixed $M_\hyper$ and $M_\local$ scales as $\bigO(J)$.

The full SwiG replacement step is then given in \cref{alg:nsswig}, where we perform $M$ total alternating pairs of hyperparameter and local sweeps. This number of repeats is exposed as a tunable hyperparameter. In theory, $M=1$ suffices to update all parameters once; in practice we use $M=5$ for more conservative mixing.

Each block update is an exact slice-sampling update targeting the corresponding full conditional under $\prior_{\thresh}$. Therefore the alternating hyperparameter update and local sweep define a Gibbs kernel that leaves $\prior_{\thresh}$ invariant.

\subsection{Complete Algorithm and Implementation}\label{subsec:implementation}
    
With a defined mechanism to perform the constrained sampling inner loop, we further embed this within a vectorized nested sampling outer loop. Nested sampling is closely related to SMC~\citep{chopin2020introduction,salomone2024unbiasedconsistentnestedsampling}, maintaining $m$ live particles approximately distributed according to the current constrained prior $\prior_{\thresh}$, with $\thresh$ increasing over iterations. These particles can be updated in parallel by deleting $k$ particles with the lowest likelihoods (thereby raising $\thresh$ to the new minimum live likelihood) and replacing them with $k$ new samples from the constrained prior, using the procedure of \cref{subsec:gibbs_theory}. The choice of $k/m$ is analogous to setting a target Effective Sample Size (ESS) in adaptive tempering SMC methods~\citep{Fearnhead2010}. 

We implement NS-SwiG by extending the Nested Slice Sampling (NSS) framework of \citet{Yallup2026}, which provides a vectorized formulation of nested sampling using Hit-and-Run~\citep{smith1984efficient} Slice Sampling~\citep{Neal2003, Rudolf_2018} for constrained updates. Tuning of the slice sampler is informed by taking a Mahalanobis-normalized covariance estimate of the previous point cloud to inform the slice proposal~\citep{Handley2015}, which we take to be block-diagonal with one block per group. The outer loop is terminated when a sufficient level of prior volume compression has occurred, such that the ratio of remaining mass in \cref{eq:evidence} is below a fixed threshold. The algorithm is implemented in JAX \citep{jax2018} within the BlackJAX framework~\citep{Cabezas2024}, and utilizes the composable transforms in JAX to parallelize MCMC moves across particles.

\subsection{Complexity and Scaling}\label{subsec:complexity}
The outer loop of particle methods adds to the runtime scaling of the methods, but provides many profitable features in return. As these algorithms generically involve compression of a prior distribution to a posterior, there is a cost that scales with the Kullback--Leibler divergence, $\dkl$, between these distributions. For hierarchical models as in \cref{eq:hierarchical_posterior}, the conditional independence structure allows decomposition of this divergence into contributions from the hyperparameters and each local block:
\begin{equation}
H \;=\; \dkl\big(\posterior(\hyper\mid \data)\,\|\,\prior(\hyper)\big)
\;+\; \mathbb{E}_{\posterior(\hyper\mid \data)}\!\left[\sum_{j=1}^J \dkl\big(\posterior(\local_j\mid \hyper,\data_j)\,\|\,\prior(\local_j\mid\hyper)\big)\right]
\;\equiv\; H_\hyper + \sum_{j=1}^J H_j.
\label{eq:info_decomp}
\end{equation}
When each group contributes $\bigO(1)$ information, $H$ scales linearly with $J$, which under standard assumptions implies that the outer iterations scale as $\bigO(J)$ \citep{Skilling2006}. This raises the overall complexity when combined with the SwiG kernel of NS-SwiG to $\bigO(J^2)$, improving over the $\bigO(J^3)$ scaling of standard joint-space nested sampling. Under the same standard assumptions the log-evidence error scales as,
\begin{equation}
    \sigma(\log \hat{\mathcal{Z}}) \approx \sqrt{H/m}\,.
\label{eq:error_scaling}
\end{equation}
These rates describe ideal Monte Carlo fluctuations; in practice, imperfect mixing can introduce additional variance that dominates the $\sqrt{H/m}$ term. This means that maintaining fixed normalizing constant precision as $J$ grows requires $m \propto J$ live particles. Similar scaling arguments apply to SMC methods~\citep{Beskos2014}, which we describe in~\cref{app:particle}.

\subsection{Comparison to MCMC Methods}

The outer-loop cost detailed in \cref{subsec:complexity} is intrinsic to methods that bridge prior to posterior, and can appear unfavourable compared to MCMC methods that target the posterior directly. There are however a number of nice features that we gain in return for this cost, namely: 
\begin{itemize}
    \item Evidence estimation. Primarily, particle methods directly estimate the evidence $\evidence$, which is crucial for model comparison and hypothesis testing tasks. MCMC methods typically require additional techniques such as bridge sampling~\citep{gelman_simulating_1998} or learned harmonic mean estimators~\citep{mcewen2023machinelearningassistedbayesian} to estimate $\evidence$, which can be computationally expensive and have less robust scaling guarantees~\citep{micaletto2025bridgesamplingdiagnostics}.
    \item Convergence diagnostics and parallelism. Particle methods typically have a natural parallelism across particles and can leverage this to assess convergence~\citep{chopin2020introduction}. Comparable diagnostics for MCMC typically require running multiple chains~\citep{Vehtari_2021}. Depending on the MCMC scheme, it can be difficult to fully recoup this cost via parallelization~\citep{pmlr-v130-hoffman21a,pmlr-v151-hoffman22a,rioudurand2023adaptivetuningmetropolisadjusted}.
    \item Tuning. Particle methods can use the point cloud of particles to implement a variety of tuning schemes adaptively. Adaptive MCMC is notoriously tricky to implement~\citep{andrieu_tutorial_2008}, and in practical settings is instead usually implemented as \emph{pretuning} with some independent warmup steps. 
\end{itemize}

It is particularly instructive to compare SwiG moves to Metropolized HMC steps. Under optimal conditions with some assumptions about the smoothness of the target, classical optimal-scaling results imply that maintaining $\bigO(1)$ acceptance probability requires a leapfrog step size $\epsilon \propto d^{-1/4}$~\citep{beskos2010optimaltuninghybridmontecarlo}. In a hierarchical model with $d \approx d_\hyper + J d_\local$, this suggests a best-case $\bigO((J d_\local)^{1/4})$ gradient evaluations per effectively independent draw. However, each gradient evaluation of the joint log density aggregates contributions from all $J$ groups, so its cost scales as $\bigO(J)$ in the number of group log-likelihood evaluations. Combining these heuristics yields a best-case computational scaling of $\bigO(J^{5/4})$ per effective posterior draw for global-gradient methods. The prior work of~\citet{Luu2024} established that for generalized linear models, there are cases when Slice-within-Gibbs can outperform HMC and vice-versa. By lifting this construction into a particle method, we target a broader class of inference problems, and explore the potential for model comparison at scale.

%% file: sections/experiments.tex
\section{Numerical Experiments}
\label{sec:experiments}

We evaluate NS-SwiG on four benchmarks, validating the method's asymptotic scaling, normalizing constant estimation, and performance on challenging Bayesian inference problems.

\paragraph{Common settings.}
Unless otherwise stated, all nested sampling runs use $m = 1000$ particles, delete $k = 50$ particles per iteration (batch deletion; see \Cref{app:glm} for an ablation), and terminate when $\log \mathcal{Z}_{\mathrm{live}} - \log \mathcal{Z} < -3$. We use $M=5$ sweeps as a default (see \Cref{app:funnel} for an ablation), with $d_\hyper$ hyperparameter slice steps and $d_\local$ local slice steps per group. We compare, where appropriate, to some baseline algorithms, with key settings detailed as follows:
\begin{itemize}
  \item NSS: standard nested slice sampling on the joint space, using $M \times d$ random-direction slice steps per replacement, where $d$ is the joint dimension. Otherwise this follows the settings of NS-SwiG for a fair comparison.
  \item SMC-HMC: adaptive-tempered SMC with HMC mutation steps~\citep{buchholz2020adaptivetuninghamiltonianmonte}, using 1000 particles with a target ESS of 0.95 and systematic resampling. Mass matrix and step size are tuned adaptively; full details are given in \Cref{app:smc_settings}.
  \item NUTS: No-U-Turn Sampler \citep{Hoffman2014} with 4 parallel chains and window adaptation warmup~\citep{Carpenter2017}. Per-experiment chain lengths and warmup settings are given in each subsection.
\end{itemize}

Evaluation cost is reported in ``full-likelihood equivalents.'' For NS-SwiG, each local update evaluates a single group log-likelihood $\loglike_j(\local_j, \hyper)$; we count each such call as $1/J$ of a full-likelihood evaluation, so that $J$ single-group calls correspond to one full evaluation. For NUTS, each leapfrog step requires one gradient evaluation and counts as one evaluation, including warmup. For SMC-HMC, evaluations are summed over all particles, HMC steps, and temperature increments. Overheads of gradient evaluations are not included as we mostly seek to validate general scaling behavior~\citep{DBLP:journals/corr/abs-1811-05031}.

Effective sample size (ESS) is computed as follows: for NUTS, we report the minimum bulk ESS across all parameters using the ArviZ library~\citep{arviz_2019}; for NS, we compute ESS from the posterior importance weights as described in~\citet{Yallup2026}; for SMC, we compute ESS from the recycled particle pool across all temperature steps~\citep{le_thu_nguyen_improving_2014}. Evaluations, runtime, and ESS are averaged over repeated runs.

Tables report $\log \hat{\mathcal{Z}}$ as the mean over 5 independent seeds $\pm$ one standard deviation (seed-to-seed), and $\hat{\sigma}$ is the average single-run internal uncertainty estimator, derived using standard recipes~\citep{Skilling2006,chopin2020introduction}. All algorithms are implemented using BlackJAX \citep{Cabezas2024}, and run on an Apple M1 Studio Max CPU. Further experimental details are provided in \Cref{app:experiments}.

\subsection{Hierarchical Gaussian Model}
\label{subsec:glm}
We first evaluate the performance and scaling of NS-SwiG on an idealized target with tractable posteriors and evidence, primarily to confirm the scaling arguments of both the number of evaluations and the normalizing constant error. We consider a hierarchical Gaussian model with $d_\hyper = 1$ hyperparameter $\hyper$ (the population mean) and $J$ groups each with $d_\local = 1$ local parameter $\local_j$, with one observation $y_j$ per group:
\begin{align}
\hyper &\sim \mathcal{N}(\mu_0, \sigma_\hyper^2), \\
\local_j \mid \hyper &\sim \mathcal{N}(\hyper, \sigma_\local^2), \\
y_j \mid \local_j &\sim \mathcal{N}(\local_j, \sigma_{\mathrm{obs}}^2).
\end{align}
Marginalizing $\local_j$ yields $y_j \mid \hyper \sim \mathcal{N}(\hyper, \tau^2)$ with $\tau^2 = \sigma_\local^2 + \sigma_{\mathrm{obs}}^2$, and marginalizing $\hyper$ then gives the analytic log-evidence.
The joint parameter dimension is $d = d_\hyper + J \cdot d_\local = 1 + J$, and the posterior information satisfies $H = \bigO(J)$.
We use $\mu_0 = 0$, $\sigma_\hyper = 10$, $\sigma_\local = 2$, and $\sigma_{\mathrm{obs}} = 1$, with data generated from $\hyper = 3$. In this model, the observation likelihood $p(y_j \mid \local_j)$ is independent of $\hyper$, so updating $\hyper$ requires recomputing only the conditional priors $\prior(\local_j \mid \hyper)$.
Although $\hyper$ updates do not require likelihood calls, we force a full recomputation during $\hyper$ updates to emulate the typical setting where the likelihood depends on hyperparameters. Evaluation counts reflect this additional cost.

\begin{figure}[t]
\centering
\subfloat[Full-likelihood equivalents: $\bigO(J)$ vs.\ $\bigO(J^2)$]{\includegraphics[width=0.32\textwidth]{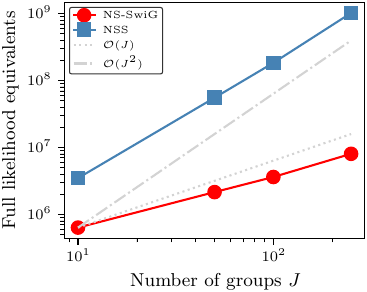}\label{fig:glm_nss_cost}}
\hfill
\subfloat[Normalized evidence error]{\includegraphics[width=0.32\textwidth]{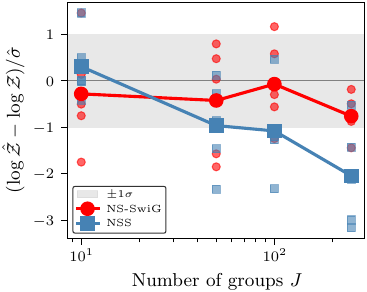}\label{fig:glm_nss_evidence}}
\hfill
\subfloat[Wall-clock runtime]{\includegraphics[width=0.32\textwidth]{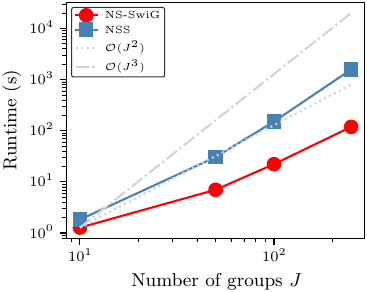}\label{fig:glm_nss_runtime}}

\vspace{0.5em}

\subfloat[Full-likelihood equivalents ($\bigO(J)$)]{\includegraphics[width=0.32\textwidth]{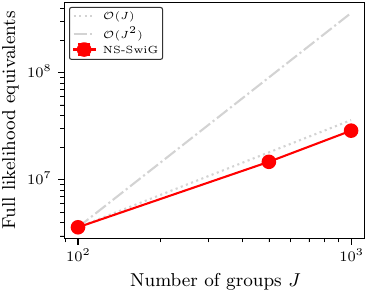}\label{fig:glm_scaling_cost}}
\hfill
\subfloat[Evidence error vs.\ $J$]{\includegraphics[width=0.32\textwidth]{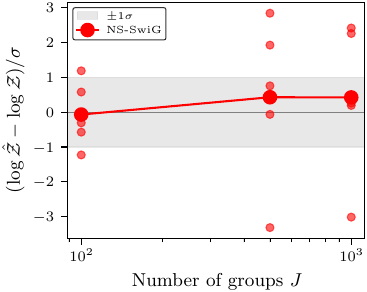}\label{fig:glm_scaling_evidence}}
\hfill
\subfloat[Wall-clock runtime vs.\ $J$]{\includegraphics[width=0.32\textwidth]{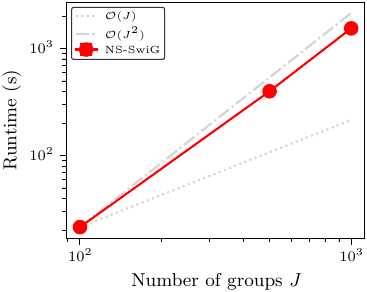}\label{fig:glm_scaling_runtime}}
\caption{Hierarchical Gaussian model scaling. \textbf{Top row (a--c):} NS-SwiG vs.\ NSS for $J \in \{10, 50, 100, 250\}$. \textbf{Bottom row (d--f):} NS-SwiG extended scaling to $J \in \{100, 500, 1000\}$.}
\label{fig:glm_scaling}
\end{figure}

First, we compare NS-SwiG against standard joint sampling in NSS for $J \in \{10, 50, 100, 250\}$ and confirm that the proposed scaling improvement from $\bigO(J^2)$ to $\bigO(J)$ in full-likelihood equivalents holds (\Cref{fig:glm_nss_cost}). We also show that the normalizing constant estimate, and internal estimate of the standard error, remain accurate and consistent with the analytic value across this range. For NSS, the estimate of $\log \mathcal{Z}$ starts to degrade at large $J$, compatible with the view that the fixed $5\times d$ repeats of slice sampling is providing insufficient mixing for multivariate slice sampling in hundreds of dimensions. The evaluation scaling translates into substantial wall-clock gains at larger $J$ (\Cref{fig:glm_nss_runtime}), though for small $J$ dispatch overhead partially masks the asymptotic behavior. Numerical results averaged over 5 repeated runs are shown in \Cref{tab:glm_nss}.

\begin{table}[t]
\centering
\small
\caption{NS-SwiG vs NSS hierarchical Gaussian model scaling comparison (results are averaged over 5 repeated runs with different random seeds).}
\label{tab:glm_nss}
\begin{tabular}{lrrrrrr}
\toprule
Method & $J$ & Runtime (s) & Evaluations & $\log \mathcal{Z}_{\text{true}}$ & $\log \hat{\mathcal{Z}}$ & $\hat{\sigma}$
  \\
\midrule
NS-SwiG & $10$ & $1.3\times10^{0}$ & $6.3\times10^{5}$ & $-25.34$ & $-25.36 \pm 0.14$ & $0.13$ \\
NSS & $10$ & $1.8\times10^{0}$ & $3.5\times10^{6}$ & & $-25.30 \pm 0.09$ & $0.13$ \\
\midrule
NS-SwiG & $50$ & $7.0\times10^{0}$ & $2.1\times10^{6}$ & $-120.64$ & $-120.76 \pm 0.29$ & $0.26$ \\
NSS & $50$ & $3.0\times10^{1}$ & $5.5\times10^{7}$ & & $-120.90 \pm 0.24$ & $0.27$ \\
\midrule
NS-SwiG & $100$ & $2.2\times10^{1}$ & $3.6\times10^{6}$ & $-228.71$ & $-228.74 \pm 0.30$ & $0.34$ \\
NSS & $100$ & $1.5\times10^{2}$ & $1.8\times10^{8}$ & & $-229.09 \pm 0.30$ & $0.34$ \\
\midrule
NS-SwiG & $250$ & $1.2\times10^{2}$ & $8.0\times10^{6}$ & $-559.76$ & $-560.19 \pm 0.26$ & $0.56$ \\
NSS & $250$ & $1.6\times10^{3}$ & $1.0\times10^{9}$ & & $-560.82 \pm 0.48$ & $0.52$ \\
\bottomrule
\end{tabular}
\end{table}

Second, we scale up to $J = 1000$ groups ($d = 1001$), dropping NSS as its runtime becomes prohibitive. \Cref{fig:glm_scaling_cost,fig:glm_scaling_evidence,fig:glm_scaling_runtime} confirm that $\bigO(J)$ scaling in full-likelihood equivalents holds, and that evidence estimates remain accurate. The observed seed-to-seed variance of $\log \hat{\mathcal{Z}}$ is slightly larger than the average internal estimator $\hat{\sigma}$, suggesting a small additional source of mixing error, but the overall estimates remain within acceptable precision even at $J = 1000$.

\begin{table}[t]
\centering
\small
\caption{NS-SwiG hierarchical Gaussian model extended scaling results (results are averaged over 5 repeated runs with different random seeds).}
\label{tab:glm_scaling}
\begin{tabular}{rrrrrrr}
\toprule
$J$ & $d$ & Runtime (s) & Evaluations & $\log \mathcal{Z}_{\text{true}}$ & $\log \hat{\mathcal{Z}}$ & $\hat{\sigma}$ \\
\midrule
$100$ & $101$ & $2.2\times10^1$ & $3.6\times10^6$ & $-228.71$ & $-228.74 \pm 0.30$ & $0.34$ \\
$500$ & $501$ & $4.0\times10^2$ & $1.5\times10^7$ & $-1097.69$ & $-1097.36 \pm 1.56$ & $0.73$ \\
$1000$ & $1001$ & $1.5\times10^3$ & $2.9\times10^7$ & $-2204.60$ & $-2204.16 \pm 1.99$ & $1.04$ \\
\bottomrule
\end{tabular}
\end{table}

\subsection{Funnel}
\label{subsec:funnel}

The funnel distribution introduced by \citet{Neal2003} encapsulates a known pathology that is encountered in many hierarchical models, namely a strong coupling between hyperparameters and local parameters that creates a funnel-shaped geometry with varying local scales. We use a version with $d_\hyper = 1$, $J = 10$, and $d_\local = 1$ (total $d = 11$):
\begin{align}
\hyper &\sim \mathcal{N}(0, \sigma^2_\hyper), \quad \sigma^2_\hyper = 9, \\
\local_j \mid \hyper &\sim \mathcal{N}(0, e^{\hyper}), \quad j = 1, \ldots, J.
\end{align}
We treat the Gaussian conditionals as per-group likelihoods $\loglike_j = \log \mathcal{N}(\local_j \mid 0, e^\hyper)$ and assign a broad uniform prior $\local_j \sim \mathrm{Uniform}(-100, 100)$ on each local parameter to ensure support covers the funnel's varying width. The unnormalized target is therefore $\prior(\hyper) \prod_j \mathcal{N}(\local_j \mid 0, e^\hyper)$ over the uniform support, and $\log \mathcal{Z}_{\text{true}} = -52.98$ is obtained by analytically marginalizing $\local_j$ within the uniform bounds.
When $\hyper$ is large, the $\local_j$ are diffuse; when $\hyper$ is small (especially negative), they concentrate tightly near zero, creating a challenging geometry for joint-space samplers. The centered parameterization is a known failure mode of gradient-based samplers such as NUTS, which struggle to explore the funnel without reparameterization~\citep{gorinova2019automaticreparameterisationprobabilisticprograms}. We therefore also test a non-centered version: $\eta_j \sim \mathcal{N}(0,1)$ with $\local_j = \eta_j \exp(\hyper/2)$, which renders the problem well-conditioned.

\begin{figure}[t]
\centering
\subfloat[NS-SwiG and NSS marginals]{\includegraphics[width=.33\textwidth]{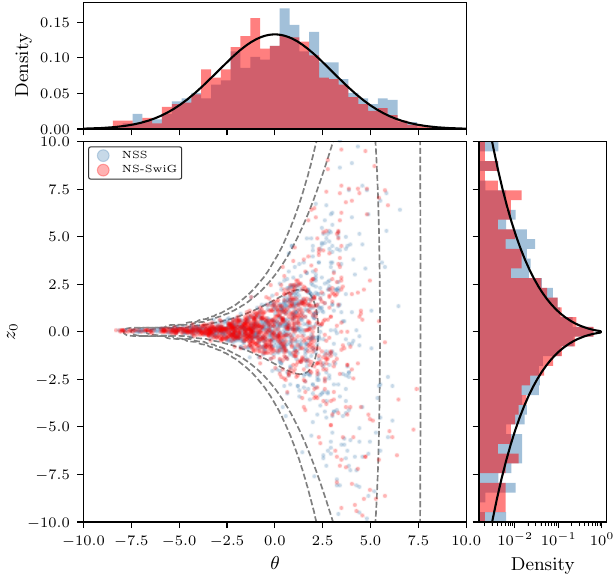}\label{fig:funnel_ns}}%
\subfloat[SMC-HMC and NUTS marginals]{\includegraphics[width=.33\textwidth]{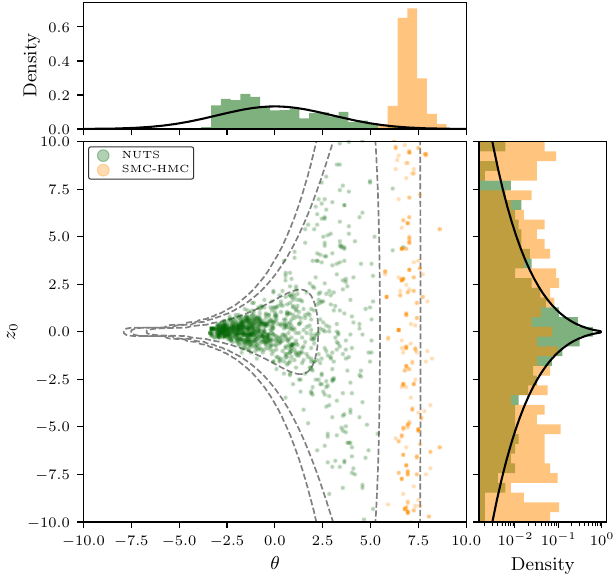}\label{fig:funnel_baselines}} \\
\subfloat[ESS per likelihood evaluation]{\includegraphics[width=.66\textwidth]{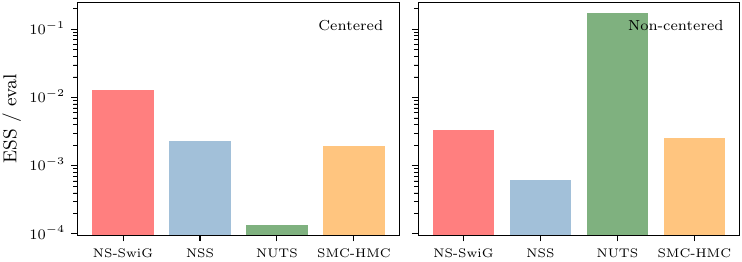}\label{fig:funnel_ess}}
\caption{Neal's funnel (centered parameterization). (a,b)~$(\hyper, \local_0)$ marginal against analytic contours (grey dashed). (c)~ESS per evaluation for both parameterizations.}
\label{fig:funnel}
\end{figure}

We compare all four algorithms on this benchmark. NUTS uses 4 chains of 2000 samples, discarding the first 500 as burn-in, with 100 steps of window adaptation. Results are shown in \Cref{tab:funnel}, with the $(\hyper, \local_0)$ marginal for the centered parameterization in \Cref{fig:funnel}.

\begin{table}[t]
\centering
\small
\caption{Neal's funnel results (results are averaged over 5 repeated runs with different random seeds). $\log \mathcal{Z}_{\text{true}} = -52.98$.}
\label{tab:funnel}
\begin{tabular}{lrrrrrr}
\toprule
Method & Runtime (s) & Evaluations & ESS & ESS/eval & $\log \hat{\mathcal{Z}}$ & $\hat{\sigma}$ \\
\midrule
NS-SwiG (C) & $5.0\times10^{0}$ & $3.8\times10^{6}$ & $49728$ & $1.3\times10^{-2}$ & $-53.03 \pm 0.10$ & $0.18$ \\
NSS (C) & $1.0\times10^{1}$ & $2.2\times10^{7}$ & $49774$ & $2.3\times10^{-3}$ & $-52.94 \pm 0.19$ & $0.18$ \\
NUTS (C) & $3.2\times10^{0}$ & $2.9\times10^{5}$ & $40$ & $1.4\times10^{-4}$ & --- & --- \\
SMC-HMC (C) & $1.0\times10^{0}$ & $8.0\times10^{5}$ & $1560$ & $2.0\times10^{-3}$ & $-56.70 \pm 0.74$ & $0.04$ \\
\midrule
NS-SwiG (NC) & $3.4\times10^{0}$ & $2.3\times10^{6}$ & $7865$ & $3.4\times10^{-3}$ & $-52.98 \pm 0.28$ & $0.20$ \\
NSS (NC) & $6.3\times10^{0}$ & $1.3\times10^{7}$ & $7926$ & $6.2\times10^{-4}$ & $-52.90 \pm 0.10$ & $0.19$ \\
NUTS (NC) & $2.6\times10^{0}$ & $5.5\times10^{4}$ & $9577$ & $1.7\times10^{-1}$ & --- & --- \\
SMC-HMC (NC) & $1.0\times10^{0}$ & $1.0\times10^{6}$ & $2602$ & $2.6\times10^{-3}$ & $-53.99 \pm 1.75$ & $0.06$ \\
\bottomrule
\end{tabular}
\end{table}

Only the nested sampling methods capture the full funnel geometry in the centered parameterization (\Cref{fig:funnel_ns}). NUTS produces numerous divergent transitions, confirming the geometry is pathological for HMC. SMC-HMC returns a biased evidence estimate with an internal uncertainty far smaller than the actual error, indicating particle degeneracy during tempering.
Under the non-centered parameterization, all methods perform well, with NUTS being most efficient (\Cref{fig:funnel_ess}). NS-SwiG maintains comparable performance to joint-space NSS while achieving the expected scaling improvement in evaluations.

\subsection{Contextual Effects Models}
\label{subsec:radon}

A contextual effects model applied to observations of radon levels across Minnesota counties \citep{gelman_data_2007} is a classic  hierarchical regression example. The dataset comprises 946 observations across $J = 85$ Minnesota counties, obtained from the \texttt{inference\_gym} benchmark suite \citep{Sountsov2020}. This model has $d_\hyper = 6$ hyperparameters $\hyper \in \mathbb{R}^6$ (county effect mean and scale, 3 regression weights, observation scale) and $J = 85$ groups each with $d_\local = 1$ local parameter (per-county random effects), yielding total dimension $d = d_\hyper + J \cdot d_\local = 91$.

We evaluate both centered and non-centered parameterizations, comparing the performance of all algorithms. NUTS uses 4 chains of 5000 samples with 1000 warmup steps, discarding the first 1000 as burn-in. To assess posterior sample quality, we compute the maximum mean discrepancy (MMD)~\citep{JMLR:v13:gretton12a} on the joint hyperparameter distribution using a Gaussian kernel with data-dependent bandwidth (see \Cref{app:mmd}), against an independent long-run non-centered NUTS reference (4 chains, 50\,000 post-warmup samples). The results are shown in \Cref{tab:radon}, with the ESS per evaluation shown in \Cref{fig:radon_ess}.

NS-SwiG yields accurate posterior marginals and evidence estimates in both parameterizations. The centered parameterization is more challenging: NS-SwiG shows higher seed-to-seed variance, but the centered estimate ($-1132.7 \pm 2.1$) is consistent with the non-centered estimates from both NS-SwiG ($-1130.0 \pm 0.1$) and SMC-HMC ($-1130.5 \pm 0.3$), suggesting the NS-SwiG centered estimate is unbiased; note that $\hat{\sigma}$ underestimates the true variance here. In contrast, SMC-HMC exhibits a ${\sim}5$ nat discrepancy in the centered case ($-1127.6 \pm 0.4$), with an internal uncertainty far smaller than the actual error, indicating particle degeneracy during tempering. SMC-HMC also shows signs of particle diversity collapse in the non-centered case: the recycled ESS (${\sim}1000$, comparable to $m$) and elevated MMD relative to the NUTS reference suggest insufficient mixing even in this better-conditioned geometry. While NUTS is more efficient for posterior sampling in the non-centered parameterization (higher ESS/eval and faster runtime), NS-SwiG remains practical and additionally provides evidence estimates.

\begin{table}[t]
\centering
\small
\caption{Radon contextual effects results (results are averaged over 5 repeated runs with different random seeds). Parenthesized MMD values indicate self-comparison of the NUTS NC reference.}
\label{tab:radon}
\begin{tabular}{lrrrrrrr}
\toprule
Method & Runtime (s) & Evaluations & ESS & ESS/eval & $\log \hat{\mathcal{Z}}$ & $\hat{\sigma}$ & MMD \\
\midrule
NS-SwiG (C) & $1.6\times10^{2}$ & $1.0\times10^{7}$ & $17686$ & $1.7\times10^{-3}$ & $-1132.71 \pm 2.12$ & $0.21$ & $2.1\times10^{-3}$ \\
NUTS (C) & $7.2\times10^{0}$ & $4.1\times10^{5}$ & $1027$ & $2.5\times10^{-3}$ & --- & --- & $1.4\times10^{-3}$ \\
SMC-HMC (C) & $3.2\times10^{1}$ & $1.6\times10^{7}$ & $1056$ & $6.5\times10^{-5}$ & $-1127.63 \pm 0.44$ & $0.06$ & $1.5\times10^{-3}$ \\
\midrule
NS-SwiG (NC) & $1.5\times10^{2}$ & $9.4\times10^{6}$ & $17012$ & $1.8\times10^{-3}$ & $-1129.99 \pm 0.06$ & $0.20$ & $1.4\times10^{-3}$ \\
NUTS (NC) & $7.3\times10^{0}$ & $4.4\times10^{5}$ & $4112$ & $9.4\times10^{-3}$ & --- & --- & ($1.2\times10^{-3}$) \\
SMC-HMC (NC) & $3.4\times10^{1}$ & $1.7\times10^{7}$ & $980$ & $5.8\times10^{-5}$ & $-1130.46 \pm 0.34$ & $0.06$ & $3.3\times10^{-3}$ \\
\bottomrule
\end{tabular}
\end{table}

\subsection{Stochastic Volatility}
\label{subsec:sv}

As a final test we consider a stochastic volatility (SV) model \citep{Kim1998} on S\&P~500, another standard benchmark from \citet{Sountsov2020} used for testing samplers at scale. SV models are often treated as a state-space model and tackled with particle filtering/SMC methods that exploit the model's sequential structure~\citep{Andrieu2010,Chopin2013}. We employ this model as a stress-test for scaling to high-dimensional joint inference, noting that specialized particle methods for models with this structure are available.

Applying the SwiG kernel requires accounting for the Markov structure of the latent variables, which breaks the conditional independence underlying the budget decomposition. We extend the implementation to maintain local budgets for each block and its neighbours, so that feasibility checks remain $\bigO(1)$ (details in \Cref{app:markov_swig}). This necessitates the centered parameterization, as the non-centered version renders the likelihood non-local, breaking the budget decomposition. Here $J$ denotes the number of time steps (returns), playing the same role as the number of groups in previous experiments. We consider $J=100$ (small) and $J=2516$ (full series). The model has $d_\hyper = 3$ hyperparameters $\hyper \in \mathbb{R}^3$ (persistence $\beta$, level $\mu$, volatility-of-volatility $\sigma$) and $J$ latent log-volatilities ($d_\local = 1$), giving total dimension $d = d_\hyper + J = \{103, 2519\}$.

\begin{figure}[t]
\centering
\subfloat[Radon model]{\includegraphics[width=0.43\textwidth]{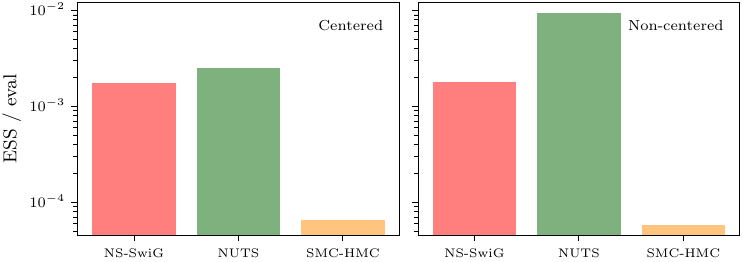}\label{fig:radon_ess}}
\hfill
\subfloat[SV small ($d=103$)]{\includegraphics[width=0.25\textwidth]{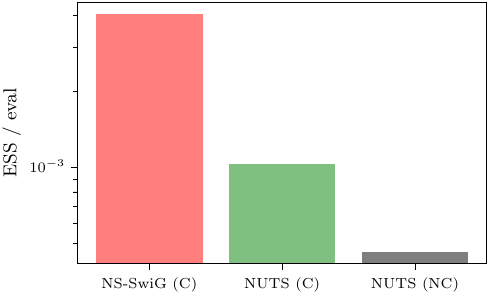}\label{fig:sv_ess_small}}
\hfill
\subfloat[SV full ($d=2519$)]{\includegraphics[width=0.25\textwidth]{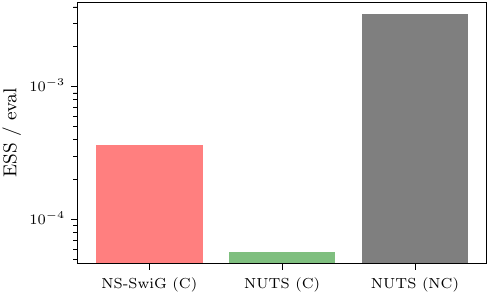}\label{fig:sv_ess_full}}
\caption{ESS per likelihood evaluation across benchmarks.}
\label{fig:radon_sv_ess}
\end{figure}

We compare NS-SwiG (centered) against NUTS in both parameterizations (4 chains, 100\,000 post-warmup samples, 10\,000 warmup steps discarded). SMC-HMC is omitted: the extreme variance in log-likelihood under the prior prevented reliable initialization of the tempering schedule, and modifying the prior to address this would compromise the point of computing evidences. MMD is computed on the joint hyperparameter distribution using the same Gaussian kernel as in \Cref{subsec:radon}. Results are shown in \Cref{tab:sv} and \Cref{fig:sv_ess_small,fig:sv_ess_full}.

For $J=100$, NUTS in the non-centered parameterization produces frequent divergent transitions, indicating that the non-centered transformation introduces problematic geometry for the AR(1) structure. The centered parameterization is better behaved, with higher ESS and no divergences. NS-SwiG produces accurate posterior marginals (\Cref{fig:app_sv_hyperparams}) and provides evidence estimates, while requiring fewer evaluations and less wall-clock time than either NUTS variant. For the full series ($J=2516$, $d=2519$), both NUTS parameterizations produce zero divergences and perform well, suggesting the additional dimensionality regularizes the geometry. All methods produce consistent MMD values (${\sim}10^{-3}$), indicating NS-SwiG recovers accurate posteriors even at this scale. This motivates further extensions to exploit Markov structure more explicitly within the NS-SwiG framework.

\begin{table}[t]
\centering
\small
\caption{Stochastic volatility (S\&P~500) results (results are averaged over 5 repeated runs with different random seeds). Parenthesized MMD values indicate self-comparison of the NUTS NC reference.}
\label{tab:sv}
\begin{tabular}{llrrrrr}
\toprule
& Method & Runtime (s) & Evaluations & ESS & MMD & $\log \hat{\mathcal{Z}}$ \\
\midrule
\multirow{3}{*}{$d=103$}
& NUTS (NC) & $1.0\times10^2$ & $13.8\times10^6$ & $6.4\times10^3$ & $(1.0\times10^{-3})$ & -- \\
& NUTS (C) & $78$ & $14.2\times10^6$ & $14.7\times10^3$ & $1.0\times10^{-3}$ & -- \\
& NS-SwiG (C) & $20$ & $2.9\times10^6$ & $11.9\times10^3$ & $1.1\times10^{-3}$ & $-573.4 \pm 0.1$ \\
\midrule
\multirow{3}{*}{$d=2519$}
& NUTS (NC) & $1.0\times10^3$ & $14.6\times10^6$ & $51.3\times10^3$ & $(1.1\times10^{-3})$ & --\\
& NUTS (C) & $1.2\times10^3$ & $35.0\times10^6$ & $2.0\times10^3$ & $8.9\times10^{-4}$ & -- \\
& NS-SwiG (C) & $4.4\times10^3$ & $30.5\times10^6$ & $11.1\times10^3$ & $1.5\times10^{-3}$ & $-10535.5 \pm 0.5$ \\
\bottomrule
\end{tabular}
\end{table}

%% file: sections/discussion.tex
\section{Discussion}
\label{sec:discussion}

The line of inquiry presented in this paper was motivated by two main prior threads of work: firstly the difficulty noted in constructing a robust, general-purpose gradient-based \emph{constrained} sampler, well summarized by \citet{kroupa2025resonances}, and secondly the retrospective observation that for models with factorizable graph structure, sampling can be made highly scalable just by exploiting this model structure~\citep{Luu2024}. We demonstrated that combining these insights, by constructing a Slice-within-Gibbs kernel that exploits factorised likelihood structure, enables nested sampling to scale to high-dimensional problems while retaining the core advantages of direct evidence estimation and robustness to challenging posterior geometries. By caching per-group likelihood contributions, the algorithm reduces the likelihood-evaluation cost per Gibbs sweep from $\mathcal{O}(J^2)$ to $\mathcal{O}(J)$ for $J$ groups, and is a step towards improving nested sampling analyses currently hitting this quadratic cost~\citep{NANOGrav:2023icp,Lovick:2025wdj}.

Empirically, we provided a comparison to multiple state-of-the-art baselines using common (and mostly conservative) settings. The experiments demonstrate that a SwiG update can yield a competitive inference kernel even at scale, with particularly strong performance in estimating normalizing constants, a notoriously difficult task at high dimension, and on challenging geometries without reparameterisation.

\paragraph{Limitations}

The main limitation of this work is the need for some kind of factorised structure in the likelihood, which is a common but not universal feature of high dimensional problems. The approach is most compelling when local blocks are low-dimensional (so that slice-based local moves mix well) and when global structure can be cached cheaply. If both the hyperparameters and each local block are high-dimensional, then the basic Slice-within-Gibbs strategy may not be the right inner kernel, and collapsing or alternative constrained proposals may be preferable.

Another practical limitation for nested sampling at high dimension is not only the cost per constrained step, but also the number of particles $m$ required to maintain accurate evidence estimates as dimensionality grows.
While NS-SwiG reduces the cost of generating constrained prior samples, and our vectorised implementation can amortise likelihood computation across groups and particles, the particle requirement can still become the dominant cost beyond $d\gtrsim 10^3$.
This motivates exploring strategies that reduce the effective dimension seen by the sampler.

More broadly, nested sampling accuracy depends on the quality of constrained samples: when the inner MCMC does not mix adequately within the likelihood contour, evidence estimates can be biased. This is a general concern for any nested sampling implementation, and motivates monitoring diagnostics such as the stability of $\log Z$ estimates with respect to the number of inner MCMC steps. Additionally, caching per-group likelihood contributions incurs $\mathcal{O}(mJ)$ memory overhead, which can become limiting on accelerators at the scales we target.

\paragraph{Extensions}
A well motivated extension that addresses the two limitations above is to adopt pseudo-marginal or more generally \emph{collapsed} approaches for subsets of parameters so that the constrained inner loop operates on a lower-dimensional representation. This is particularly natural in light of observations such as the full SV model, where increasing the latent dimension actually regularizes the problem, rendering it better approximated by conditional Gaussian assumptions that would be natural in a collapsed sampler.

Another natural case we do not explore is strongly heterogeneous compute costs across blocks, as was noted for the classic implementation of \emph{fast-slow}. It is natural to consider the blocked Gibbs approach as applicable to such problems, however, without explicit factorization of the likelihood, correlations between nominally ``fast'' and ``slow'' components could lead to more challenging dynamics to explore.

Finally, we emphasise that the strong performance of NS-SwiG on non-linear, non-Gaussian examples (including centered parameterisations) is suggestive of more than just an alternative to reparameterisation.
In many potential application domains---especially those involving physical forward models---there may be no clear \emph{a priori} transformation that reliably improves conditioning. Further validating the performance of nested sampling at scale on such problems is an interesting direction for future work, and may be a domain where NS-SwiG is particularly compelling.




%% file: sections/conclusion.tex
\section{Conclusion}
\label{sec:conclusion}

We presented NS-SwiG, a nested sampling algorithm for hierarchical Bayesian inference that reduces the cost of each live-point replacement from $\mathcal{O}(J^2)$ to $\mathcal{O}(J)$ (in full-likelihood-evaluation equivalents), where $J$ is the number of groups. The key idea is a decomposition of the likelihood-threshold constraint that allows each local block update to check feasibility in $\mathcal{O}(1)$ rather than $\mathcal{O}(J)$ time. Across benchmarks, we validate the resulting scaling and show that NS-SwiG enables robust, gradient-free posterior sampling and evidence estimation in regimes where gradient-based samplers can be unreliable.

The strength of this approach comes primarily from exploiting known factorization structure in the likelihood, which is not guaranteed in all inference problems. Notable counterexamples that require a fully joint treatment of the parameters include sampling molecular configurations~\citep{partay_nested_2014}, an area that has seen active work on neural-network-based learning of full joint configurations (e.g.,~\citet{akhoundsadegh2024iterated}). Even in such settings, there is often lower-dimensional structure that can be factorized out, such as collective variables in molecular systems~\citep{frohlking_learning_2025}. Framing the problem as discovering such low-dimensional structure when it is not known a priori, rather than inheriting the unfavorable scaling of fully joint sampling, is an interesting direction for future work.

Particularly within astrophysics, where nested sampling is already extremely popular, scaling challenges of existing inference pipelines abound. As astronomical catalogs continue to grow---with current and future gravitational-wave detectors, large-scale surveys of galaxies and galaxy clusters---methods that scale efficiently with catalog size while preserving access to the evidence become increasingly important. NS-SwiG provides both posterior samples and Bayesian evidence in a single run, making nested sampling practical for large-scale hierarchical problems where scalable inference and rigorous model comparison are both required.

%% file: appendix.tex
\newpage
\appendix

\section{Complete NS-SwiG Algorithm}
\label{app:algorithm}

Algorithm~\ref{alg:ns_outer} presents the outer nested sampling loop with batch deletion, and Algorithm~\ref{alg:nsswig} presents the NS-SwiG replacement kernel used as the constrained update step.

\begin{algorithm}[t]
\caption{Nested Sampling with Batch Deletion}
\label{alg:ns_outer}
\small
\begin{algorithmic}[1]
\Require Prior $\prior(\vartheta)$, likelihood $L(\vartheta)$, $m$ live particles, batch size $k$, termination criterion $\epsilon$
\Ensure Dead particles $\{\vartheta_i, \thresh_i\}$, evidence estimate $\hat{\mathcal{Z}}$
\State Draw $m$ live particles $\{\vartheta^{(i)}\}_{i=1}^m$ from $\prior(\vartheta)$
\State Evaluate $\loglike^{(i)} = \log L(\vartheta^{(i)})$ for all $i$
\While{$\log \mathcal{Z}_{\mathrm{live}} - \log \hat{\mathcal{Z}} > \epsilon$} \Comment{$\mathcal{Z}_{\mathrm{live}} = X_t \cdot \max_i L^{(i)}$}
    \State \textbf{Reweight:} identify the $k$ particles with lowest $\loglike^{(i)}$; set $\thresh$ to the $k$-th lowest
    \State Record these $k$ particles as dead samples; update $\hat{\mathcal{Z}}$ via NS quadrature~\citep{Skilling2006}
    \State \textbf{Resample:} select $k$ parents uniformly from the surviving $m - k$ live particles
    \State \textbf{Mutate:} for each parent, apply constrained update (Alg.~\ref{alg:nsswig}) \Comment{$\parallel$ over $k$}
    \State \textbf{Replace:} insert the $k$ new particles into the live set
\EndWhile
\State Append remaining $m$ live particles as final dead samples
\State \Return dead particles, $\hat{\mathcal{Z}}$
\end{algorithmic}
\end{algorithm}

\begin{algorithm}[t]
\caption{SwiG: Slice-within-Gibbs Constrained Update}
\label{alg:nsswig}
\small
\begin{algorithmic}[1]
\Require Live points $\{(\hyper^{(i)}, \{\local_j^{(i)}\}, S^{(i)}, \{\loglike_j^{(i)}\})\}_{i=1}^m$, threshold $\thresh$
\Ensure Updated live point satisfying $S > \thresh$
\State Select point $i$; set $(\hyper, \{\local_j\}, S, \{\loglike_j\}) \leftarrow (\hyper^{(i)}, \{\local_j^{(i)}\}, S^{(i)}, \{\loglike_j^{(i)}\})$
\For{$s = 1, \ldots, M$} \Comment{$M$ Gibbs sweeps}
    \Statex \hspace{\algorithmicindent}\rule{0.85\columnwidth}{0.4pt}
    \Statex \hspace{\algorithmicindent}\textbf{OUTER:} Update $\hyper \mid \{\local_j\}$; target $\prior(\hyper) \prod_j \prior(\local_j|\hyper) \cdot \mathbf{1}[\sum_j \loglike_j > \thresh]$
    \Statex \hspace{\algorithmicindent}\rule{0.85\columnwidth}{0.4pt}
    \State $\hyper \leftarrow \text{SliceSample}\bigl(\hyper;\ f(\cdot) = \log \prior(\cdot) + \textstyle\sum_j \log \prior(\local_j | \cdot),\ c(\cdot) = \sum_j \loglike_j(\local_j, \cdot) > \thresh\bigr)$
    \State $\loglike_j \leftarrow \loglike_j(\local_j, \hyper)$ for all $j$;\ \ $S \leftarrow \sum_j \loglike_j$ \Comment{$\bigO(J)$}
    \Statex \hspace{\algorithmicindent}\rule{0.85\columnwidth}{0.4pt}
    \Statex \hspace{\algorithmicindent}\textbf{INNER:} Sweep $\local_j \mid \hyper, \local_{-j}$; target $\prior(\local_j|\hyper) \cdot \mathbf{1}[\loglike_j > \budget_j]$
    \Statex \hspace{\algorithmicindent}\rule{0.85\columnwidth}{0.4pt}
    \For{$k = 1, \ldots, J$}
        \State $\budget_k \leftarrow \thresh - S + \loglike_k$ \Comment{$\bigO(1)$}
        \State $\local_k \leftarrow \text{SliceSample}\bigl(\local_k;\ f(\cdot) = \log \prior(\cdot | \hyper),\ c(\cdot) = \loglike_k(\cdot, \hyper) > \budget_k\bigr)$
        \State $\loglike_k \leftarrow \loglike_k(\local_k, \hyper)$;\ \ $S \leftarrow S - \loglike_k^{\mathrm{old}} + \loglike_k$ \Comment{$\bigO(1)$}
    \EndFor
\EndFor
\State \Return $(\hyper, \{\local_j\}, S, \{\loglike_j\})$
\end{algorithmic}
\end{algorithm}

\section{Connection to Particle Methods}
\label{app:particle}
Nested sampling maintains a particle population, and can be formalised within a broader family of particle methods through the lens of SMC~\citep{salomone2024unbiasedconsistentnestedsampling}. We have focused on defining connections to how the nested sampling algorithm is used in practice, primarily in astrophysics, namely as targeting a static target and using artificial bridging distributions to improve MCMC exploration and provide estimates of Bayesian evidences. Given the depth of the SMC literature in comparison to the nested sampling literature, we reflect on some possible connections that a SwiG kernel can make to SMC samplers, and how the scaling results of \cref{subsec:complexity} have classical analogues for SMC samplers.

\subsection{NS-SwiG as resample-move in joint space}
\label{app:resample_move}

At level $\thresh$, nested sampling maintains particles
$\{(\hyper^{(i)},\{\local_j^{(i)}\})\}_{i=1}^m$ targeting
\begin{equation}
\prior_{\thresh}(\hyper,\{\local_j\}) \propto \prior(\hyper)\prod_{j=1}^J \prior(\local_j\mid \hyper)\,
\mathbf{1}\!\left[\sum_{j=1}^J \loglike_j(\local_j,\hyper)>\thresh\right],
\end{equation}
with $\thresh$ increasing via order statistics and evidence accumulated by the nested sampling quadrature rule.
This ``population + MCMC rejuvenation'' structure closely parallels SMC samplers and resample-move schemes \citep[e.g.,][]{gilks_following_2001,chopin_sequential_2002,DelMoral2006}, but with nested sampling specific level selection and weight/evidence bookkeeping.

Replacing a live point applies a Markov kernel $K_{\thresh}$ that leaves $\prior_{\thresh}$ invariant.
In NS-SwiG, $K_{\thresh}$ is a blocked Gibbs sweep with slice sampling \citep{Neal2003}.
The hierarchical conditional independence structure implies that each local update depends only on $(\hyper,\local_j)$ and the global likelihood constraint; budget caching makes the constraint check $\bigO(1)$ in $J$ per local update, yielding $\bigO(J)$ per sweep (in full-likelihood equivalents), as established in \cref{subsec:complexity}.

\subsection{Rare-event view and global rejuvenation via particle Gibbs}
\label{app:rare_event_pg}

Fix $\hyper$ and $\thresh$.
The local conditional target is
\begin{equation}
\prior_{\thresh}(\{\local_j\}\mid \hyper)\propto \Big[\prod_{j=1}^J \prior(\local_j\mid \hyper)\Big]\,
\mathbf{1}\Big\{\sum_{j=1}^J \loglike_j(\local_j,\hyper)>\thresh\Big\}.
\end{equation}
Ordering groups as $j=1,\dots,J$ and defining $S_j=\sum_{i=1}^j \loglike_i(\local_i,\hyper)$ yields a sequential view: drawing $\local_j\sim \prior(\local_j\mid \hyper)$ evolves
$S_{j-1}\mapsto S_j=S_{j-1}+\loglike_j$.
The constraint $S_J>\thresh$ is a terminal rare event, which explains why naive sequential proposals over $j$ degenerate: particles remain equally weighted until the final indicator, where almost all weights are zero when the event is rare.

If block updates struggle to mix well, this can be motivation to consider mixing occasional \emph{global} rejuvenation moves.
A standard exact option is particle Gibbs (conditional SMC) \citep{Andrieu2010}, used as an MCMC kernel on the full trajectory $\local_{1:J}$ by conditioning an internal particle system on the current configuration $\local_{1:J}^{\mathrm{ref}}$ (guaranteeing at least one surviving path).
In difficult regimes, particle Gibbs with ancestor sampling \citep{lindsten2014particlegibbsancestorsampling} can reduce path degeneracy and stickiness.
Because groups are exchangeable, randomizing the group order can reduce artifacts.

\subsection{Pseudo-marginal directions}
\label{app:streaming_ns2}

A large part of the SMC literature is motivated by state-space and latent-variable models in which the marginal likelihood $p(y\mid \theta)$ is analytically intractable but can be estimated unbiasedly using particle filters. This observation underpins particle MCMC and related constructions, where unbiased likelihood estimators are embedded within higher-level Monte Carlo schemes, and it is exploited explicitly by SMC$^2$ \citep{Chopin2013}, which combines an \emph{outer} SMC sampler over static parameters with an \emph{inner} SMC algorithm that delivers unbiased likelihood estimates for each parameter particle (useful in both batch and streaming regimes). Such state-space formulations have not found wide traction in astrophysics, but the underlying pseudo-marginal principle is broadly applicable. Relatedly, pseudo-marginal slice sampling \citep{pmlr-v51-murray16} shows that slice sampling can be carried out using only \emph{unbiased} (typically nonnegative) likelihood estimators, by augmenting the state with the randomness used to form the estimate. Since nested sampling enforces a likelihood-threshold constraint at level $\thresh$, it can also be viewed through an auxiliary-variable lens closely related to slice sampling. This suggests future work on pseudo-marginal variants of nested sampling, potentially extending evidence estimation to models with intractable likelihoods but available unbiased estimators.

\subsection{SMC error scaling}
\label{app:smc_scaling}

The scaling results of \cref{subsec:complexity} have classical analogues for SMC samplers.
High-dimensional stability analyses show that, for additive log-likelihood targets satisfying appropriate regularity conditions, the number of tempering steps must increase with dimension to prevent weight degeneracy: specifically, temperature increments should scale as $\Delta\beta = \bigO(1/d)$, yielding $T$ scaling linearly with $d$ \citep{Beskos2014}.
For hierarchical models where each group contributes $\bigO(1)$ effective dimension and $d$ scales linearly with $J$, this implies $\bigO(J)$ tempering steps are required.

Central limit theorems for Feynman--Kac models establish that the normalizing constant estimator satisfies $\sigma(\log \hat{\mathcal{Z}}) = \bigO(m^{-1/2})$ for a fixed number of steps \citep{DelMoral2004,Chopin2004}.
When the number of steps grows as $\bigO(J)$, each contributing $\bigO(1)$ variance, the total asymptotic variance scales linearly with $J$, yielding $\sigma(\log \hat{\mathcal{Z}}) = \bigO(\sqrt{J/m})$.
Consequently, maintaining fixed normalizing-constant precision as $J$ grows requires $m \propto J$ particles, matching the nested sampling conclusion of \cref{subsec:complexity}.
These rates describe ideal Monte Carlo fluctuations; in practice, imperfect mixing of rejuvenation kernels can introduce additional variance.

\subsection{Slice-within-Gibbs within SMC}
\label{app:smc_swig}

The SwiG kernel could equally be used as a rejuvenation kernel within tempering-based SMC, exploiting the same conditional independence structure to achieve $\bigO(J)$ sweeps.
For a tempered joint target $\pi_t(\hyper,\local_{1:J}) \propto \prior(\hyper)\prod_{j=1}^J \prior(\local_j\mid\hyper)\exp\{\beta_t\loglike_j\}$, the $\local_k$-conditional has the form $\prior(\local_k\mid\hyper)\exp\{\beta_t\loglike_k\}$, which becomes sharply peaked as $\beta_t\to 1$, requiring narrower slice widths and more shrinkage steps.
In contrast, NS-SwiG targets the truncated prior $\prior(\local_k\mid\hyper)\cdot\mathbf{1}[\loglike_k > \budget_k]$, which remains relatively flat within the feasible region.
This makes NS-SwiG potentially easier to tune in practice, but we leave exploration of a SMC-SwiG variant to future work.

\section{Additional Experimental Details}
\label{app:experiments}

This appendix provides supplementary plots and details for each experiment in Section~\ref{sec:experiments}.

\subsection{SMC-HMC Settings}
\label{app:smc_settings}

The SMC-HMC baseline uses the following tuning strategy. The mass matrix assumes a diagonal structure, estimated from the covariance of the previous iteration's particle cloud. The step size is tuned using the previous iteration's mean expected squared jump distance (ESJD). Trajectory lengths are set to 5 for the hierarchical Gaussian and funnel benchmarks, and 20 for the contextual effects benchmark. SMC-HMC is omitted from the stochastic volatility benchmark due to initialization difficulties (see \Cref{subsec:sv}).

\subsection{MMD Computation}
\label{app:mmd}

The maximum mean discrepancy (MMD) is computed on the joint hyperparameter distribution (not individual 1D marginals) using a Gaussian kernel. The bandwidth is set to the average pairwise $L^2$ distance between all samples in the combined reference and test sets. We use 1000 samples from each method, subsampled uniformly at random, and repeat this procedure 5 times. The average across this set, repeated over multiple reseeded runs where available is then used in the tables. The reference is always the non-centered NUTS posterior.

\subsection{Hierarchical Gaussian Model}
\label{app:glm}

\begin{figure}[t]
\centering
\subfloat[Posterior samples for $(\hyper, \local_0)$ at $J = 250$ ($d = 251$). NS-SwiG (red), $1\sigma$ and $2\sigma$ contours (black), NSS (blue).]{\includegraphics[width=0.48\textwidth]{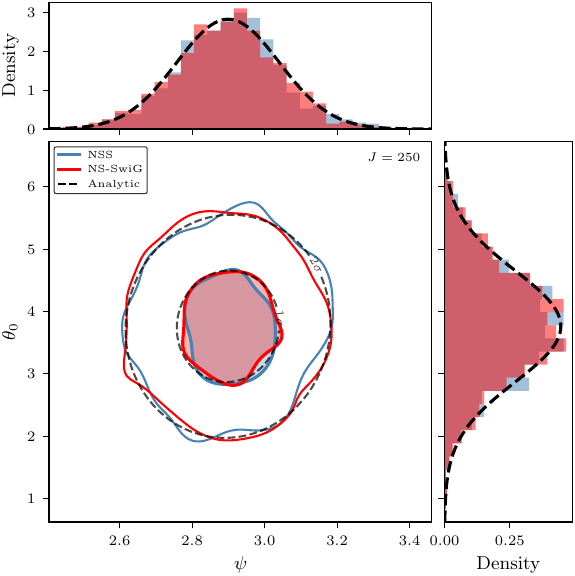}\label{fig:app_glm_nss}}
\hfill
\subfloat[Posterior samples at $J = 1000$ ($d = 1001$). NS-SwiG (red) matches the analytic contours, demonstrating effective mixing in high dimensions.]{\includegraphics[width=0.48\textwidth]{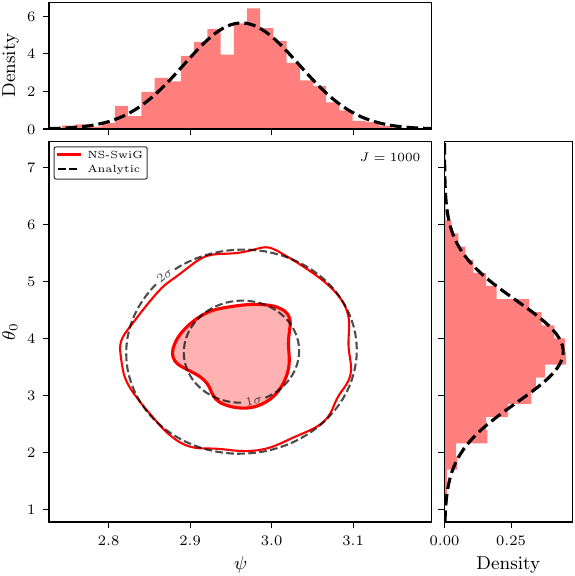}\label{fig:app_glm_scaling}}
\caption{Hierarchical Gaussian model: 2D marginal posteriors for $(\hyper, \local_0)$.}
\label{fig:app_glm_marginals}
\end{figure}

\Cref{fig:app_glm_marginals} shows the 2D marginal posteriors for $(\hyper, \local_0)$ at $J = 250$ (NS-SwiG vs.\ NSS comparison) and $J = 1000$ (large-scale benchmark). NS-SwiG matches the analytic posterior contours in both cases, whereas NSS shows visible degradation at $J = 250$ due to mixing difficulties in the high-dimensional joint space.

To complement the scaling results in \Cref{subsec:glm}, we ablate the batch deletion size $k$ on the hierarchical Gaussian at $J=100$. \Cref{tab:glm_num_delete} shows that increasing $k$ from 1 to 50 reduces runtime by over an order of magnitude with negligible impact on evidence accuracy or ESS, validating batch deletion as a practical acceleration.

\begin{table}[t]
\centering
\small
\caption{NS-SwiG hierarchical Gaussian ($d=101$, $\log \mathcal{Z}_{\text{true}} = -228.70$) with varying batch deletion size $k$ (results are averaged over 3 repeated runs).}
\label{tab:glm_num_delete}
\begin{tabular}{rrrrrr}
\toprule
$k$ & Runtime (s) & ESS & Likelihood evals & $\log \hat{\mathcal{Z}}$ & $\hat{\sigma}$ \\
\midrule
$1$ & $814$ & $18.0\times10^3$ & $5.0\times10^6$ & $-229.0 \pm 0.2$ & $0.29$ \\
$10$ & $170$ & $16.4\times10^3$ & $4.9\times10^6$ & $-228.7 \pm 0.2$ & $0.29$ \\
$25$ & $73$ & $16.6\times10^3$ & $4.9\times10^6$ & $-229.1 \pm 0.3$ & $0.29$ \\
$50$ & $48$ & $17.8\times10^3$ & $4.8\times10^6$ & $-228.4 \pm 0.3$ & $0.30$ \\
\bottomrule
\end{tabular}
\end{table}

\subsection{Neal's Funnel}
\label{app:funnel}

To complement the funnel results in \Cref{subsec:funnel}, we ablate the number of Gibbs sweeps $M$ per replacement on the 10D funnel. \Cref{tab:funnel_sweeps} shows that evidence accuracy is stable across all values of $M$, while evaluation cost scales linearly. Even $M=1$ produces accurate results on this problem, though more sweeps may be needed for harder geometries.

\begin{table}[t]
\centering
\small
\caption{NS-SwiG 10D funnel ($d=11$, $\log \mathcal{Z}_{\text{true}} = -52.98$) with varying number of Gibbs sweeps $M$ (results are averaged over 3 repeated runs).}
\label{tab:funnel_sweeps}
\begin{tabular}{rrrrrr}
\toprule
$M$ & Runtime (s) & ESS & Likelihood evals & $\log \hat{\mathcal{Z}}$ & $\hat{\sigma}$ \\
\midrule
$1$ & $2.0$ & $20.5\times10^3$ & $0.6\times10^5$ & $-53.0 \pm 0.2$ & $0.18$ \\
$2$ & $2.4$ & $18.0\times10^3$ & $1.3\times10^6$ & $-53.0 \pm 0.2$ & $0.18$ \\
$3$ & $3.1$ & $19.7\times10^3$ & $1.9\times10^6$ & $-53.0 \pm 0.1$ & $0.19$ \\
$4$ & $3.8$ & $18.1\times10^3$ & $2.5\times10^6$ & $-53.2 \pm 0.2$ & $0.18$ \\
$5$ & $4.4$ & $15.0\times10^3$ & $3.1\times10^6$ & $-52.9 \pm 0.1$ & $0.18$ \\
\bottomrule
\end{tabular}
\end{table}

\subsection{Contextual Effects Models}
\label{app:radon}

\begin{figure}[t]
\centering
\includegraphics[width=\textwidth]{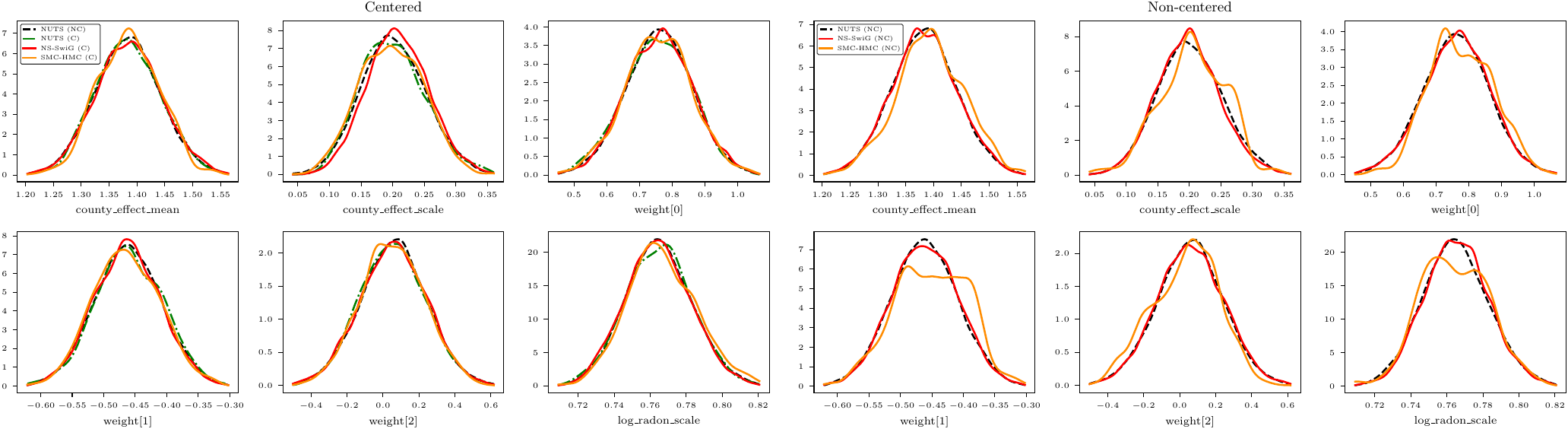}
\caption{Radon contextual effects model: hyperparameter marginals for centered (left three panels) and non-centered (right three panels) parameterizations. NUTS non-centered (black dashed) serves as reference. NS-SwiG (red), NUTS centered (green), and SMC-HMC (orange) all produce consistent marginals.}
\label{fig:app_radon_hyperparams}
\end{figure}

\Cref{fig:app_radon_hyperparams} shows hyperparameter marginals for both centered and non-centered parameterizations, compared against the NUTS non-centered reference.

\subsection{Markov SwiG Kernel}
\label{app:markov_swig}

The iid SwiG kernel (Algorithm~\ref{alg:nsswig}) exploits the fact that, conditional on $\hyper$, the local variables $\local_j$ are independent under the prior. For models with Markov latent structure (e.g.\ AR(1) stochastic volatility), the prior factorizes as
\begin{equation}
\prior(\local_{0:T-1}\mid\hyper) = \prior(\local_0\mid\hyper)\prod_{t=1}^{T-1}\prior(\local_t\mid\local_{t-1},\hyper),
\end{equation}
while the likelihood remains site-local: $L = \prod_{t=0}^{T-1} L_t(\local_t,\hyper)$. Since the likelihood factorizes over sites, the budget decomposition from Algorithm~\ref{alg:nsswig} still applies:
\begin{equation}
\budget_t = \thresh - \sum_{s\neq t}\loglike_s,
\end{equation}
and the running sum $S = \sum_t \loglike_t$ is updated incrementally after each site, maintaining $\bigO(1)$ feasibility checks. Updating $\local_t$ requires the conditional prior given its Markov blanket $\{\local_{t-1},\local_{t+1}\}$:
\begin{equation}
\log\prior(\local_t \mid \text{blanket}, \hyper) =
\begin{cases}
\log\prior(\local_0\mid\hyper) + \log\prior(\local_1\mid\local_0,\hyper), & t=0,\\[4pt]
\log\prior(\local_t\mid\local_{t-1},\hyper) + \log\prior(\local_{t+1}\mid\local_t,\hyper), & 0<t<T{-}1,\\[4pt]
\log\prior(\local_{T-1}\mid\local_{T-2},\hyper), & t=T{-}1.
\end{cases}
\end{equation}
Each site update therefore targets $\prior(\local_t\mid\text{blanket},\hyper)\cdot\mathbf{1}[\loglike_t > \budget_t]$, which is the product of at most two transition densities and a likelihood indicator. Sites are updated in sequential order $t=0,1,\ldots,T{-}1$. This sequential dependence is unavoidable given the Markov prior, but each site update costs $\bigO(1)$ likelihood evaluations and $\bigO(1)$ prior evaluations, giving $\bigO(T)$ per sweep as in the iid case.

In a standard AR(1) non-centering, one writes $\local_t = \mu + \sigma\,\eta_t$ where $\eta_t \sim \mathcal{N}(\beta\,\eta_{t-1},\,1)$, so that the innovations $\eta_t$ absorb the dependence on $\hyper = (\mu,\sigma,\beta)$. However, the observation likelihood $L_t(y_t \mid \local_t) = L_t(y_t \mid \mu + \sigma\,\eta_t)$ now depends on the hyperparameters $\mu$ and $\sigma$ at \emph{every} site. This means updating a single innovation $\eta_t$ changes $\local_t$, which couples to $\hyper$ in the likelihood, breaking the per-site factorization: $\loglike_t$ is no longer a function of $(\eta_t)$ alone, and the budget $\budget_t = \thresh - \sum_{s\neq t}\loglike_s$ cannot be evaluated without recomputing all likelihood terms that share $\hyper$. This is why the SV experiments use centered parameterization exclusively.

\subsection{Stochastic Volatility}
\label{app:sv}

\begin{figure}[t]
\centering
\subfloat[Small model ($J=100$, $d=103$).]{\includegraphics[width=\textwidth]{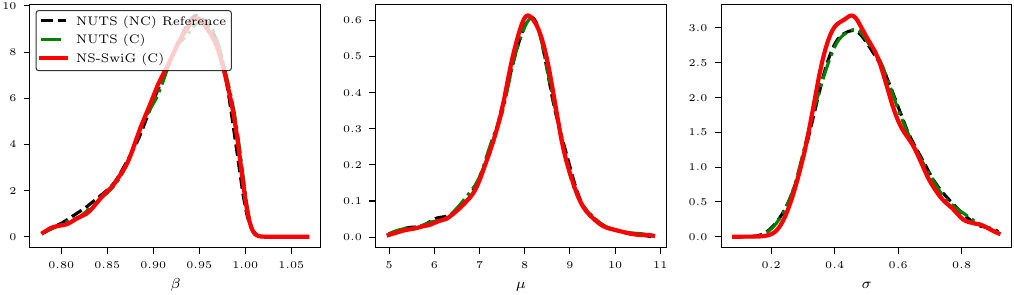}\label{fig:app_sv_small}}

\vspace{0.5em}

\subfloat[Full model ($J=2516$, $d=2519$).]{\includegraphics[width=\textwidth]{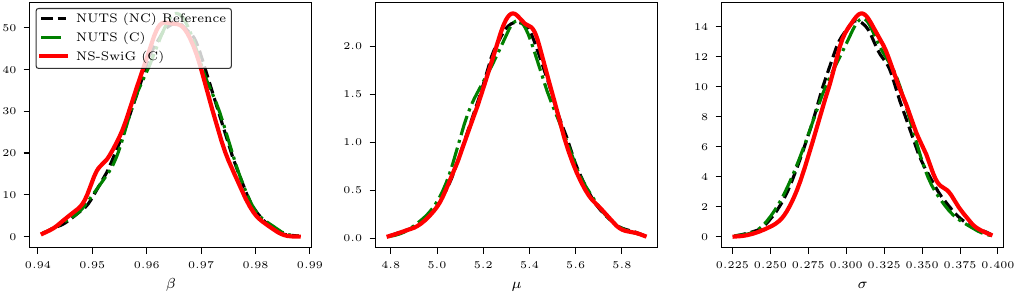}\label{fig:app_sv_full}}
\caption{SV model hyperparameter marginals. NUTS non-centered (black dashed) serves as reference. NS-SwiG Markov (red, centered) and NUTS centered (green) both recover the reference posteriors.}
\label{fig:app_sv_hyperparams}
\end{figure}

\Cref{fig:app_sv_hyperparams} shows hyperparameter marginals for both the small ($J=100$, $d=103$) and full ($J=2516$, $d=2519$) SV models, comparing NS-SwiG Markov (centered) against NUTS (centered and non-centered).

